\documentclass[conference]{IEEEtran}
\IEEEoverridecommandlockouts
\usepackage{cite}
\usepackage{amsmath,amssymb,amsfonts}
\usepackage{algorithmic}
\usepackage{graphicx}
\usepackage{comment}
\usepackage{textcomp}
\usepackage{xcolor}
\usepackage{tikz}
\usepackage{amsfonts}
\usepackage{amsmath}
\usepackage{xurl}
\usepackage{caption}  
\usepackage{tabularx}
\usepackage[numbers]{natbib}
\usepackage{bbm}
\DeclareMathOperator{\E}{\mathbb{E}}
\DeclareMathOperator{\I}{\mathbbm{1}}
\usetikzlibrary{automata,arrows,positioning,calc}
\usetikzlibrary{fit}
\usepackage{pgfplots}
\usepackage{subcaption}
\usepackage[numbers]{natbib}
\usepackage{float}

\def\BibTeX{{\rm B\kern-.05em{\sc i\kern-.025em b}\kern-.08em
    T\kern-.1667em\lower.7ex\hbox{E}\kern-.125emX}}
\begin{document}

\title{Analytic Framework for Estimating Memory Cost\\}

\author{\IEEEauthorblockN{Anirudh Shankar, Avhishek Chatterjee, and Anjan Chakravorty}
\IEEEauthorblockA{Indian Institute of Technology Madras, India}}

\maketitle

\begin{abstract}
As artificial intelligence (AI) models quickly spread and become more advanced, they are requiring an ever-increasing amount of data and compute capability, leading to a significant energy cost. Training and inference of AI models including the large language models (LLMs) and deep neural networks (DNNs) are contributing to a large carbon footprint owing to the massive amount of memory they consume in data centers. In this article, we present a generalized framework that quantifies these energy costs incurred to the environment. This framework provides a foundational quantification of AI's ecological footprint, facilitating the development of sustainable architectural strategies for future models. 
\end{abstract}


\section{Introduction}
Rapid development in artificial intelligence (AI), machine learning (ML) and large language models (LLMs) has resulted in unprecedented demand for data with an energy cost that is rising at an alarming rate \cite{patterson2021carbonemissionslargeneural}, \cite{9942310}. Data centers generate enormous amounts of heat and need millions of gallons of water to cool down. These facts reveal growing concerns about the long-term sustainability and environmental impacts of memory and computing technologies closely associated with AI. 

Traditional charge-based memories require frequent data refreshes due to their lower retention time, contributing to a significant energy cost. Emerging memory technologies such as stochastic magnetic tunnel junctions (sMTJs) have demonstrated promising capabilities for reduced energy consumption and high retention times, making them suitable for energy-efficient applications like probabilistic or p-bit computing \cite{borders2019integer}, \cite{chowdhury2023full}. However, their manufacturing process that involves specific materials like ferromagnetic metals and insulating oxides, does carry an environmental footprint related to resource extraction, chemical usage, energy consumption during fabrication, and potential challenges of e-waste if not managed sustainably \cite{9235579}. Thus, there appears to exist a trade-off between the material- and process-related costs to the environment and the operating energy-related costs to the environment. This necessitates the requirement of a cost framework that can be used to compare across devices, architectures, and topologies of different memories, taking into account all the potential costs involved. 

Through this article, we propose a generalized \textbf{COST} framework that takes into account all the environmental hazards, including e-waste generation, operating energy costs, and other related costs to contrast different memory devices across materials, architectures, and topologies.  In \cite{shankar2024stochasticanalysisretentiontime}, we presented the concept of calculating retention time to estimate the energy cost involved in memory usage. Here, we will build up on this earlier work to propose the generalized \textbf{COST} framework in Section \ref{sec:cost} and compare the framework across several memory topologies presented in \cite{shankar2024stochasticanalysisretentiontime} in Section \ref{sec:sec2}. Finally, we discuss the limitations of this framework in \ref{sec:conc}. 

% \textcolor{red}{Request to add more in introduction with respect to energy cost/ rising power demands. Most of the survey has been done in Paper 1 \cite{shankar2024stochasticanalysisretentiontime}}

% {\AC It would be good to recollect a bit about the memory topologies from the previous paper. The readers should be able to read this paper as a stand-alone paper by just trusting the results from the previous paper.}

\subsection{Our Contributions}

\textcolor{black}{Through this article, we aim to introduce a framework that could potentially lay the groundwork for estimating the energy efficiency of AI systems. By using the notion of a generalized retention time defined in \cite{shankar2024stochasticanalysisretentiontime}, we analyze the energy efficiency of some simple coupled networks. This paper can prove to be a critical missing connection between low-level device physics and algorithmic hardware efficiency. }

\textcolor{black}{Recent studies, such as \cite{aquino2025towards}, have attempted to address the requirement of an all-unifying energy framework by establishing standardized energy consumption indices to compare various deep learning architectures and GPU hardware. Software-based energy estimators such as CodeCarbon \cite{benoit_courty_2024_11171501}, CarbonTracker \cite{Anthony2020CarbontrackerTA} and Green's Algorithm \cite{https://doi.org/10.1002/advs.202100707} provide high-level metrics for tracking real time carbon emissions. However, these works abstract out the underlying hardware and device physics by assuming it as a black box. Consequently, these models overlook the fundamental laws of physics that drive power dissipation at the device level. }

\textcolor{black}{Therefore, in order to address the rising concern over the carbon footprint of AI, optimizing the next generation of computing systems at just the software level is not sufficient. To attain true sustainability, a bottom-up approach must be followed, optimizing each layer of the stack, starting from the device physics that govern the system. This necessitates the requirement of a framework that can analytically compute the energy efficiency of a system taking into account the energy cost of bit retention and switching energy and contrast it with already existing systems. By identifying the specific physical parameter that governs the most energy loss, a system designer can make modifications at the top level to make the system much more energy efficient. For example, the energy required to refresh DRAM cells is a major contributor of energy loss in a microprocessor \cite{6237001}. The present research attempts to provide the necessary benchmarking scheme in building high-performance AI systems that are inherently aligned with reducing global carbon footprint.}

\subsubsection*{Building on the contributions of \cite{shankar2024stochasticanalysisretentiontime}} In \cite{shankar2024stochasticanalysisretentiontime}, a mathematical framework was developed for accurately modeling the energy requirement for refreshing different memory architectures. There we discussed the physics of memory deterioration in a thermal bath, i.e., system noise using non-equilibrium statistical mechanics, in particular Glauber dynamics. A careful Markov analysis of the physics-based model offered a quantitative understanding of the rate of memory refreshing in different fault-tolerant memory architectures in terms of different parameters of the technology and architectures. This finally led to a quantitative evaluation of the energy requirement. 

The quantitative results arising from a statistical physics driven Markov analysis in \cite{shankar2024stochasticanalysisretentiontime} provides the foundation for the overall cost analysis in the present paper. A major cost of memory is their energy requirement, which is taken directly from \cite{shankar2024stochasticanalysisretentiontime} and subsequently combined with other costs. Naturally, these different costs are influenced by a common set of architectural and technology parameters. In the present paper, we develop a framework to analyze the overall cost by factoring in the contributions of the different sources. By considering a few example architectures and technology choices, we show the efficacy of our framework in optimizing the overall cost of any memory by plugging in the right technology parameters.

\section{The \textbf{COST} Framework}
\label{sec:cost}

Based on that framework elaborated in \cite{shankar2024stochasticanalysisretentiontime}, here we develop a systematic approach for comparing the effective cost of storage in terms of energy efficiency across different memory architectures, topologies, and materials. Although the present paper discusses the cost structure in terms of magnetic memories, one can generalize it for other kinds of memories by suitably mapping the cost components to address the specific cases.

For any engineering system, total cost has two main components, namely the one-time cost, sometimes referred to as the capex, and the recurring cost, sometimes referred to as the opex. The one-time cost captures the cost of manufacturing and setting-up of an engineering system amortized over the lifetime of the system. Recurring cost captures the cost per unit time for running the system as desired.

For a memory, the one-time cost is primarily dictated by the choice of the materials, their properties, and the corresponding fabrication technologies.

\subsection{One-Time Costs}
In this section, we shall look at the one-time costs that must be incurred to use the block of memory. Note that these costs once paid suffice for the entire lifetime of the block. Thus, in a way, these costs are also one-time costs per unit time. 
\begin{itemize}
        \item [(1)] Material Cost : $C_M$ \\ 
        The material cost is a fixed one-time cost to the environment at the time of purchase or extraction of material. This cost could be thought of as a factor that accounts for the environmental impact due to the material extraction from nature, manufacturing and disposing-off of the memory block after usage. This cost indirectly measures the electronic-waste and carbon footprint generated in the process. From the perspective of the magnetic memories \cite{kimata2019magnetic}, \cite{chen2014anomalous}, \cite{zhu2003new}, this cost is a function of the material and processing techniques used in their fabrication. However, a crucial component of the processing cost enabling or preserving the coupling strength between the dipoles is counted separately under the coupling related processing cost.
        \item [(2)] Coupling Related Processing Cost : $C(s_f)$ \\ 
         Introducing or preserving the magnetic anisotropy in the memory material to enhance the retention time \cite{zhang2025stress} can be thought of as an example of processing conditions leading to an increase in the overall one-time cost. Essentially, to ensure retentivity, some sort of coupling must be introduced between the dipoles. This results in an additional cost that depends on the choice of the underlying purification and fabrication process that ensures sufficient coupling between the dipoles ($s_f$). We shall refer to this cost as $C(s_f, C_M)$. Since the chosen process depends on the material involved, this cost also depends on $C_M$. For the sake of simplicity of notation, we shall refer to this cost as $C(s_f)$. That is, 
        \begin{equation}
            C(s_f) = g(s_f, C_M)
        \end{equation} 
        where $g(.)$ is a function that varies from material to material and the strength of coupling. We define $g(s_{f}, C_{M})$ as a monotonically increasing function; higher coupling strengths and superior material quality inherently necessitate more complex, and thus more costly, fabrication processes. Empirically, we shall just assume that a simple power law can predict how $C(s_f)$ would depend on both $s_f$ and $C_M$ as
        \begin{equation}
            C(s_f) = k s_f^m C_M^n.
            \label{csfx}
        \end{equation} 
        Our framework is robust to the variation of $m$ and $n$ and can easily be factored in based on the material and coupling strength. This is another fixed one-time cost to the environment and has the same consequences as the material cost. This coupling cost could be thought of as an additional energy cost that needs to be paid for. To ensure that the dipoles interact with each other, a cost that accounts for the interaction energy between these dipoles needs to be accounted for. This interaction energy is usually quadratic in $s_f$ \cite{griffiths2023introduction}. However, we have taken an arbitrary polynomial index $m$ that could account for the material dependence of the dipole interactions \cite{wang2018magnetic}. Since the number of dipoles is limited to either one or three in our analysis, the energy cost of the interaction of the magnetic field with the dipoles is negligible. Here, $k$ is a constant of appropriate dimensions. \\
        $C(s_f)$ also depends on fabrication conditions \cite{mcevoy2015materials}, \cite{misra2023probabilistic}. Thus, it makes sense to account for the dependence of $C(s_f)$ on processing parameters as well. Since our earlier cost $C_M$ accounts for all process dependencies and waste generations, the coupling cost has a dependence on $C_M$ as shown in Equation \eqref{csf}. \\
        However, for the sake of demonstrating the efficacy of our framework, we demonstrate results assuming $m=1$ and $n=1$ for a simplistic analysis that produces
        \begin{equation}
            C(s_f) = k s_f C_M.
            \label{csf}
        \end{equation} 
    \end{itemize}

\subsection{Recurring Costs}
In addition to one-time costs, one must pay recurring costs (i) to maintain a constant magnetic ($H$) field and (ii) to periodically refresh the data after a retention time ($\tau$). These costs are explained below. 
\begin{itemize} 
        \item [(1)] Cost of maintaining constant $H$ field : $C(H)$ \\ 
        The cost of maintaining a constant $H$ field is given by $C(H)$ \cite{griffiths2023introduction} and is a continuous energy cost to the environment. This variable cost to the environment is accounted for in the form of energy consumed over the lifetime of the block and is thus an energy-per-unit cost expressed as
        \begin{equation*}
            C(H)=\frac{H^2}{2 \mu}
        \end{equation*}
        where $\mu$ is the magnetic permeability of the magnetic medium used in the memory.
        \item [(2)] Replenishment Cost : $C_R$ \\ 
        The cost of recharging the memory block with the data originally stored is given by the replenishment cost ($C_R$) \cite{9495814}. In our framework, $C_R$ is the replenishment cost for \textit{each dipole}. However, since the data are recharged only after the retention time \cite{shankar2024stochasticanalysisretentiontime}, it is meaningful to introduce the effective replenishment cost given by 
        \begin{equation*}
            C_{eff}(R)=\frac{C(R)}{\E[T]} =  \frac{C(R)}{\tau}
        \end{equation*}
        %$C_{eff}(R)=\frac{C(R)}{\E[T]} =  \frac{C(R)}{\tau}$ 
        %\begin{equation*}
        %    C_{eff}(R) = \frac{C(R)}{\E[T]} =  \frac{C(R)}{\tau}
        %\end{equation*}
        where $\E[T]$ is the retention time $\tau$. $C_{eff}(R)$ is a consequence of the Renewal Reward Theorem \cite{paper9} and is another energy-per-unit-time that needs to be added into our cost framework. 
    \end{itemize}
    The net cost per unit time or just net cost for simplicity (given by $C$) to the environment is thus the sum total of all the above mentioned fundamental costs and is calculated as \\
    \begin{center}
        Net Cost = Material Cost + Cost of maintaining constant H field  + Coupling Cost + Replenishment Cost
    \end{center}
    \begin{equation}
        C = C(M, H, s_f, R) = C_M + C(H) + C(s_f) + C_{eff}(R).
        \label{CostEqn}
    \end{equation}
    The costs involved in the systems presented in our earlier work \cite{shankar2024stochasticanalysisretentiontime} are computed and compared against each other below.
    \begin{itemize}
        \item [(1)] The isolated single dipole case is the easiest to handle. The costs involved in this case are the material cost and the replenishment cost. Due to the absence of external H field and coupling, the costs attributed due to these terms in accordance with Equation \eqref{CostEqn} vanish in this case. Substituting the value of retention time obtained in \cite{shankar2024stochasticanalysisretentiontime}, we compute the net cost for the isolated single dipole case given by Equation \eqref{C1}. 
        \begin{equation}
            C_1 = C_M + \frac{C(R)}{2}
            \label{C1}
        \end{equation}
        \item [(2)] For a single dipole in an external H field, in addition to material cost and replenishment cost, an additional cost of maintaining a constant H field $C(H)$ needs to be paid for. Plugging in the value of retention time obtained in \cite{shankar2024stochasticanalysisretentiontime}, the net cost for a single dipole in an external H field is given by Equation \eqref{C2}
        \begin{equation}
            C_2 = C_M + \frac{C(R)}{1 + e^{2 \beta H}} + \frac{H^2}{2 \mu}
            \label{C2}
        \end{equation}
        \item [(3)] The net cost for the three uncoupled dipole system, in the absence of any external H field, has contributions only from the material cost and the replenishment cost and is given by Equation \eqref{C3}
        \begin{equation}
            C_3 = 3 C_M + 3\frac{C(R)}{6}
            \label{C3}
        \end{equation}
        Since there are three dipoles that need to be paid for at the time of purchasing, and three dipoles need to be recharged at the time of recharging, the extra factor of 3 appears in the cost formulation here.
        \item [(4)] For the three uncoupled dipole system in the presence of an external $H$ field, the net cost as in Equation \eqref{C2} has contributions from three terms - the material cost, the cost to maintain a constant external H field and the replenishment cost. 
        \begin{equation}
            C_4 = 3 C_M + 3\frac{C(R)}{\tau_4} + \frac{H^2}{2 \mu}
            \label{C4}
        \end{equation}
        where $\tau_4$ represents the retention time in this scenario \cite{shankar2024stochasticanalysisretentiontime}. 
        The factor of 3 appears due to the same reasons as described in Equation \eqref{C3}.
        \item [(5)] In the case of coupling along a line, in addition to the terms in Equation \eqref{C4}, an additional coupling cost needs to be accounted for. Since there are only two couplings in the case of a linear topology, the factor of 2 appears in Equation \eqref{C5}
        \begin{equation*}
            C_5 = 3C_M + 3\frac{C(R)}{\tau_5} + \frac{H^2}{2 \mu} + 2C(s_f)
        \end{equation*}
        but $C(s_f)= ks_fC_M$ as described in Section \ref{sec:cost}
        \begin{equation}
            C_5 = (3+2s_f)C_M + 3\frac{C(R)}{\tau_5} + \frac{H^2}{2 \mu} 
            \label{C5}
        \end{equation}
        where $\tau_5$ is the retention time in this scenario \cite{shankar2024stochasticanalysisretentiontime}. 
        \item [(6)] Following along the same lines as Equation \eqref{C5}, for the case of coupling along a triangle, an additional contribution of $C(s_f)$ needs to be added due to the introduction of another nearest neighbour interaction. The net cost in this case is given by Equation \eqref{C6}.
        \begin{equation}
            C_6 = (3+3s_f)C_M + 3\frac{C(R)}{\tau_6} + \frac{H^2}{2 \mu} 
            \label{C6}
        \end{equation}
        where $\tau_6$ is the retention time in this scenario \cite{shankar2024stochasticanalysisretentiontime}. 
    \end{itemize}
\section{Comparison across Toplogies}
\label{sec:sec2}
Since the net cost $C$ described in Section \ref{sec:cost} is a direct measure of energy usage, lower value of net cost would imply lesser degradation of the environment. In this section, we try to contrast the differences across topologies described in Section \ref{sec:cost} in scenarios labelled from 1 through 6. 

\subsubsection{Single Dipole Case}
\label{SDC}
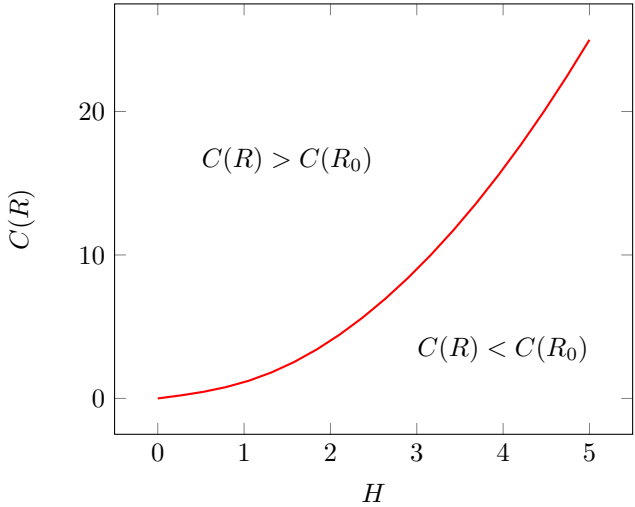
\begin{figure}[H]
    \centering
    \begin{tikzpicture}
            \begin{axis}
            [name=plot, xlabel={$H$}, ylabel={$C(R)$}]
            \addplot[red, thick] table{./Latest_Data_1.txt};
            \node [above] at (150,150) {$C(R)>C(R_0)$};
            \node [below] at (400,050) {$C(R)<C(R_0)$};
             \end{axis}
        \end{tikzpicture}
    \caption{Cost Analysis for a Single Dipole}
    \label{fig:g7}
\end{figure}

For $H>$0, it is evident that the retention time of the system with applied external H field has a higher retention time than the case without application of external $H$ field. However, the net benefit of an external H field depends on whether the resulting energy savings exceed the overhead of maintaining the field. \\
For lower energy consumption in the case of application of an external $H$ field we require constraint (\ref{Cons1}) as a consequence of our framework described in the earlier section.  
\begin{equation}
    C_2 < C_1
    \label{Cons1}
\end{equation}
From Equations \eqref{C1} and \eqref{C2}, we arrive at the following inequality following constraint \eqref{Cons1}. 
\begin{equation*}
    C_M + \frac{C(R)}{1 + e^{2 \beta H}} + \frac{H^2}{2 \mu} < C_M + \frac{C(R)}{2}
\end{equation*}
which results in 
\begin{equation}
    C(R) > \frac{H^2}{\mu} \frac{1-e^{-2 \beta H}}{1+e^{-2 \beta H}}
    \label{Cons1obt}
\end{equation}
We shall represent the right side of the constraint using the symbol $C(R_0)$ for the ease of representation. For this scenario, we have
\begin{equation*}
    C(R_0) = \frac{H^2}{\mu} \frac{1-e^{-2 \beta H}}{1+e^{-2 \beta H}}
\end{equation*}
To interpret the results, we shall consider normalizing the permeability of the material ($\mu$ = 1) and set $\beta$ = 1. Figure \ref{fig:g7} shows the plot corresponding to \ref{Cons1obt}. 
This suggests that if the replenishment cost, $C(R)$, is greater than a certain value (given by $C(R_0)$) for a particular external $H$ field, application of an external $H$ field is more energy efficient than the case without the application of an external $H$ field.  \\

Intuitively, this can be understood as follows. When the applied external $H$ field is positive, the retention time is enhanced. When the replenishment cost associated with the recharging of data is high, the frequency of recharge in the absence of external $H$ field is high resulting in a higher contribution to the effective cost. The contribution from the cost to maintain a constant $H$ field is lower than the high replenishment cost owing to frequent recharge in this case, thus, resulting in the conclusion that application of an external $H$ field is more energy efficient. 

\subsubsection{Three Uncoupled Dipoles}
Consider the isolated three dipole case in the absence and presence of an external H field having net costs given by Equations \eqref{C3} and \eqref{C4} respectively. As it was shown that for $H$>0, the retention time is enhanced significantly in the case of application of an external $H$ field in \cite{shankar2024stochasticanalysisretentiontime}, we now investigate if the application of an external $H$ field is energy efficient. This requires constraint (given by Eqn. \eqref{cons1x}) to be true. 
\begin{equation}
    C_4 < C_3
    \label{cons1x}
\end{equation}
From Equations \eqref{C3} and \eqref{C4}, we simplify \eqref{cons1x} as 
\begin{equation*}
       3 C_M + 3\frac{C(R)}{\tau_4} + \frac{H^2}{2 \mu} < 3 C_M + 3\frac{C(R)}{6}
\end{equation*}
This results in a simplified constraint 
\begin{equation}
    C(R) > \frac{H^2}{\mu} \frac{\tau_4}{\tau_4-6}
\end{equation}
where $\tau_4$ is a function of $H$ obtained in \cite{shankar2024stochasticanalysisretentiontime}. The critical replenishment cost $C(R_0)$ for this case is given by 
\begin{equation*}
    C(R_0) = \frac{H^2}{\mu} \frac{\tau_4}{\tau_4-6}
\end{equation*}
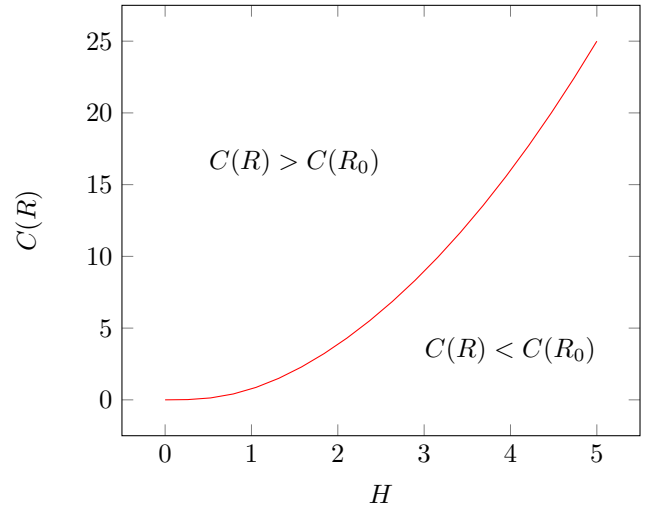
\begin{figure}[H]
    \centering
        \begin{tikzpicture}
            \begin{axis}
            [name=plot, xlabel={$H$}, ylabel={$C(R)$}]
            \addplot[red] table{./Latest_Data_2.txt};
            \node [above] at (150,150) {$C(R)>C(R_0)$};
            \node [below] at (400,050) {$C(R)<C(R_0)$};
             \end{axis}
        \end{tikzpicture}
    \caption{Three Dipole Analysis}
    \label{fig:g5}
\end{figure}

Figure \ref{fig:g5} shows the plot corresponding to the analysis of the three isolated dipoles case. As suggested by the 1-dipole case, using an external $H$ field is efficient when the replenishment cost is higher than a certain threshold. It is interesting to note that the value of critical replenishment cost ($C(R_0)$) is almost the same for the three uncoupled dipole case and the single dipole case. \\

This can be understood from the following argument. Since the replenishment cost $C(R)$ is given to \textit{each} of the dipoles in the three dipoles case, effectively, the replenishment is done to single dipoles and thus, the critical replenishment cost is almost the same. Thus, in principle, one can exploit the advantage of a high retention time at the same energy efficiency by introducing redundancy in the system as evident from this example. 

\subsubsection{Three Coupled Dipoles: Line vs Triangle}
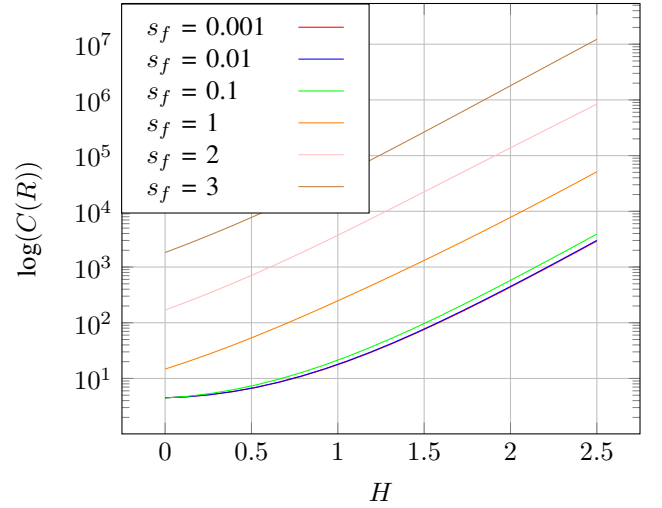
\begin{figure}[H]
    \centering
    \begin{center}
        \begin{tikzpicture}
            \begin{axis}
            [name=plot, ymode = log, xlabel={$H$}, ylabel={log$(C(R))$},legend style={at={(0.5,-0.2)},anchor=north},
      grid=major]
            \addplot[red] table{./sf1.txt};\label{$001$}
            \addplot[blue] table{./sf2.txt};\label{$01$}
            \addplot[green] table{./sf3.txt};\label{$1$}
            \addplot[orange] table{./sf4.txt};\label{$1x$}
            \addplot[pink] table{./sf5.txt};\label{$2x$}
            \addplot[brown] table{./sf6.txt};\label{$3x$}
            \end{axis}
             \node[anchor=north east ,fill=white,draw=black] (legend) at ($(plot.north)-(1.5 mm, 0 mm)$) {\begin{tabular}{l l}
        $s_f$ = 0.001 & \ref{$001$}  \\ 
        $s_f$ = 0.01 & \ref{$01$} \\
        $s_f$ = 0.1 & \ref{$1$}  \\ 
        $s_f$ = 1 & \ref{$1x$} \\
        $s_f$ = 2 & \ref{$2x$}  \\ 
        $s_f$ = 3 & \ref{$3x$}  \\ 
    \end{tabular} };
        \end{tikzpicture}
    \end{center}
    \caption{Cost as a function of H (for $m=1$)}
    \label{fig:g1}
\end{figure}

In \cite{shankar2024stochasticanalysisretentiontime}, we concluded that for the same conditions of $H$ and $s_f$, coupling along a triangle yields a much higher retention time than coupling along a line. Now, we will answer the question on whether coupling along a triangle is more energy efficient in comparison to coupling along a line. \\
For energy efficient coupling along a triangle, we require constraint \eqref{cons1xy} to be true from Section \ref{sec:cost}. 
\begin{equation}
    C_6<C_5
    \label{cons1xy}
\end{equation}
This manifests in the following relation 
\begin{equation*}
    (3+3s_f)C_M + 3\frac{C(R)}{\tau_6} + \frac{H^2}{2 \mu} < (3+2s_f)C_M + 3\frac{C(R)}{\tau_5} + \frac{H^2}{2 \mu} 
\end{equation*}
Under the assumption of using the same material and external $H$ field, we arrive at constraint \eqref{coupcos} for energy efficient usage. Note that though there is no explicit $H$ dependence in the inequality, the $H$ dependence is captured in $\tau_5$ and $\tau_6$ since they are a function of both $s_f$ and $H$. 
\begin{equation}
    C(R)>\frac{1}{3}s_fC_M\frac{\tau_5\tau_6}{\tau_6-\tau_5}
    \label{coupcos}
\end{equation}
Figure \ref{fig:g1} shows the set of curves obtained for constraint \eqref{coupcos} for several $s_f$ and $H$ values. For any curve, the region above the curve represents the condition $C(R)>C(R_0)$ = $\frac{1}{3}s_fC_M\frac{\tau_5\tau_6}{\tau_6-\tau_5}$. If the replenishment cost $C(R)$ lies in this region, it is energy efficient to couple the dipoles in a triangle rather than a linear array at that particular $s_f$ and $H$ value. \\

As the external $H$ field increases, the critical replenishment cost required $C(R_0)$ rises exponentially. Similarly, as the coupling factor $s_f$ increases, there is an exponential increase in the critical replenishment cost $C(R_0)$ required for energy efficiency of a triangular topology over a linear array. \\

Since the retention time exponentially increases with increase in $s_f$ \cite{shankar2024stochasticanalysisretentiontime}, the frequency of recharge is lowered significantly leading to lesser contribution from the replenishment cost term on both sides - the triangular topology and the linear array. However, since $s_f$ increases, the coupling cost increases, thus, resulting in the requirement of a higher critical replenishment cost ($C(R_0)$) to effectively nullify this increase in coupling cost for increase in retention time. Generalizing the framework to arbitrary exponents $m$ and $n$ (Equation \ref{csfx}) allows for a more nuanced exploration of device physics and fabrication economics. Adjusting these parameters, the model reveals that the critical replenishment cost $C(R_0)$—the threshold at which advanced architectures become beneficial—is highly sensitive to the scaling of coupling and material quality. Specifically, as these exponents increase, the "barrier to entry" for complex topologies like triangular coupling rises exponentially. This insight suggests a counterintuitive path for sustainable AI: by optimizing systems to operate effectively with lower coupling strengths and less refined materials, designers can significantly reduce the "one-time" ecological footprint while maintaining competitive energy efficiency. Essentially, this mathematical flexibility encourages the development of hardware that minimizes the total environmental impact rather than ensuring maximum performance at an unsustainable energy cost.

\section{Conclusion and Discussion}
\label{sec:conc}
In this paper, we have proposed a generic \textbf{COST} framework that can be used to analyze energy efficiency of different memory blocks across architectures, topologies and materials. This introductory framework provides a benchmarking scheme for comparing different memory devices by contrasting their energy efficiency. This paper builds on our previous work \cite{shankar2024stochasticanalysisretentiontime} on mathematical modeling of energy consumption in memory using non-equilibrium statistical mechanics and Markovian analysis. The framework proposed here can be adapted to any memory architecture and technology by choosing the corresponding parametric dependence.

This paper, building on \cite{shankar2024stochasticanalysisretentiontime}, offers a physics-driven and bottom-up framework for cost analysis of a memory. In the current era of large scale AI, this is likely to help in reducing carbon emission. Though this is a promising beginning, there are certain aspects of the work that remain open for the future. In an industrial setting, we are limited by the number of times a memory block can be written to. This limit, termed endurance, introduces a correction to our generalized cost function proposed in Section \ref{sec:cost}. This correction needs to be accounted for in our \textbf{COST} formulation, which is an important direction for future work. 

\bibliographystyle{unsrtnat}
\bibliography{references}
\end{document}